\documentclass[12pt]{iopart}

\usepackage{amsthm}
\usepackage{amsmath}
\usepackage{amsfonts}
\usepackage{amssymb}

\usepackage{graphicx}
\usepackage{dcolumn}
\usepackage{bm}
\usepackage{graphics}
\usepackage{mathtools}
\usepackage{bbm} 

\usepackage{caption}
\usepackage{subcaption}

\usepackage{float}

\usepackage{hyperref}
\hypersetup{
    colorlinks,%
    citecolor=blue,%
    linkcolor=blue,%
    urlcolor=blue
}

\newcommand{\s}{\sigma}

\newcommand{\veck}{\mathbf{k}}

\begin{document}



\title{Robustness of RKKY interactions across a Weyl node-annihilation transition}

\author{João V. F. Alves  $^1$,  Joelson F. Silva $^2$, Luis G. G. V. Dias da Silva $^1$}
\ead{luisdias@usp.br}

\address{$^1$ Instituto de Física, Universidade de São Paulo,Rua do Matão 1371,05508-090,São Paulo, SP,Brazil }

\address{$^2$ Gleb Wataghin Institute of Physics, The University of Campinas (Unicamp), 13083-859 Campinas, SP, Brazil}

\begin{abstract}
Weyl semimetals (WSMs), with their unique topological properties and distinct electronic structure, exhibit intriguing properties when either time-reversal or inversion symmetries are broken. In this work, we consider the Ruderman-Kittel-Kasuya-Yosida (RKKY) interaction between magnetic impurities in time-reversal symmetry-breaking WSMs.  We derive  analytical expressions for the full RKKY exchange tensor in arbitrary two-band spinful lattice systems. 
Our approach reveals both Heisenberg, anisotropic Ising and Dzyaloshinsky-Moriya terms, which can be calculated by energy-integrating  real-space Green's functions across the entire Brillouin zone, with the band edge acting as a natural energy cutoff.  We apply this framework to study a two-band tight-binding model for a time-reversal symmetry-breaking WSM that interpolates between a Weyl phase with well-separated chiral nodes and a quadratic band-touching semimetal phase. Remarkably, the spatial profile, magnitude, and anisotropic tensor structure of the exchange couplings remain persistent across the node-annihilation transition. This topological robustness reveals that short- and intermediate-range RKKY interactions are mediated by the global, Brillouin-zone-integrated quantum metric of the full valence band rather than being strictly dictated by local low-energy Berry curvature monopoles. 
These findings demonstrate the necessity of full-band tight-binding formulations when predicting real-space magnetic interactions, providing key insights for electric-field tuning of magnetic anisotropy and constructing realistic models of heavy-fermion and Weyl-Kondo semimetals.
\end{abstract}

\submitto{\JPCM}

\maketitle

\section{Introduction}


Weyl semimetals (WSMs) \cite{Wan:Phys.Rev.B:205101:2011,Burkov:Phys.Rev.Lett.:107:127205:2011} have emerged as captivating materials due to their exotic topological properties and unique electronic structure characterized by low-energy relativistic Dirac-like cones~\cite{weyl_review,RevModPhys.90.015001,RevModPhys.93.025002}. 
 WSWs are tree dimensional topological systems whose the breaking of time-reversal or inversion symmetries leads to a separation of Dirac-like cones in the spectrum in these materials, leading to the formation of pairs of chiral \emph{Weyl nodes} located at different points in the ${\bf k}$-space, 
where the energy dispersion is linear and chiral. 

The appearance of pairs of Weyl nodes in the spectrum has important consequences. First, these nodes are topologically protected in the sense that they harbor topologically protected chiral charges and hence, remain gapless under the presence of small perturbations such as disorder, 
leading to intriguing, graphene-like, transport phenomena \cite{Kundu:NewJournalofPhysics:013037:2023}. Secondly, a clear signature of the separation of the Weyl nodes is the presence of ``Fermi arcs" (projections of a continuous sequence of gapless states connecting the Weyl nodes) on the surface of the WSM \cite{Wan:Phys.Rev.B:205101:2011,Burkov:Phys.Rev.Lett.:107:127205:2011},
which can be experimentally probed by ARPES  in materials such as TaAs \cite{Xu:Science:613--617:2015,Yang:NaturePhysics:728:2015,Lv:Phys.Rev.X:031013:2015} and NbAs \cite{Xu:NaturePhysics:748:2015}.

Moreover, the interplay between strong electronic correlations, spin-orbit coupling, and band topology 
is most prominently exemplified by the emergence of \emph{Weyl-Kondo semimetals} \cite{Lai_PNAS_9397_2018,PhysRevB.101.075138,Chen2022,PhysRevB.109.035153,Dzsaber2021} in heavy-fermion systems such as $\text{Ce}_3\text{Bi}_4\text{Pd}_3$ \cite{Lai_PNAS_9397_2018}. In these materials, dense arrays of localized $f$-electron magnetic moments interact with itinerant conduction electrons and, according to the extended Doniach picture, this setting is governed by a fundamental competition: local Kondo screening attempts to quench local moments into non-magnetic spin singlets, while the conduction-electron-mediated Ruderman, Kittel, Kasuya, and Yosida (RKKY) interaction~\cite{PhysRev.96.99,10.1143/PTP.16.45,PhysRev.106.893} favors long-range magnetic ordering among the local moments. When Kondo hybridization dominates under appropriate crystal symmetries, heavy quasiparticles undergo extreme mass renormalization, generating emergent Weyl nodes tightly pinned near the Fermi energy with drastically reduced Fermi velocities \cite{Lai_PNAS_9397_2018,PhysRevB.101.075138,Dzsaber2021}.


The RKKY interaction in Weyl semimetals \cite{Chang:Phys.Rev.B:241103:2015,hosseini,Sun:JPhysCondensMatter:435306:2017} is a complex phenomenon influenced by various factors. Both internode process and the unique three-dimensional spin-momentum locking are relevant to RKKY-type interactions in WSMs, resulting in both Heisenberg and Ising terms, and an additional Dzyaloshinsky-Moriya term in the absence of inversion symmetry \cite{Chang:Phys.Rev.B:241103:2015,hosseini}. 
Additionally, the RKKY interaction in three-dimensional tilted Dirac and Weyl semimetals can lead to a modulation of the oscillation frequencies of range functions as the main influence of tilt on the RKKY Hamiltonian \cite{Wang:PhysicsLettersA:126327:2020}. These studies collectively underscore the complexity and richness of the RKKY interaction in Weyl semimetals. 


Within the Weyl-Kondo framework, the spatial profile and tensor anisotropy of the RKKY exchange play a decisive role in stabilizing or perturbing the topological heavy-fermion ground state. Because the RKKY interaction tensor is explicitly mediated by the host conduction states, non-trivial spin-momentum locking imparts anisotropic Heisenberg, Ising, and Dzyaloshinskii–Moriya (DM) components into the effective exchange integral. Quantifying these exchange parameters beyond linearized low-energy approximations is essential for constructing realistic Kondo-lattice Hamiltonians, mapping quantum phase boundaries, and understanding how magnetic correlations evolve when the underlying conduction spectrum undergoes topological transitions or Fermi-level shifts. Precise knowledge of full-band tight-binding RKKY exchange tensors thus provides critical input for modeling the fate of local moments and quantum criticality in candidate Weyl-Kondo semimetals.

In this paper, we investigate the RKKY interaction in a time-reversal-symmetry-breaking lattice model that smoothly interpolates between a Weyl semimetal phase with well-separated chiral nodes and a quadratic band-touching semimetal phase  formed upon node annihilation. By employing a full two-band tight-binding formulation, our approach accounts for the full Brillouin zone and high-energy valence states, thereby avoiding the integration artifacts and artificial ultraviolet regulators required in low-energy linearized continuum models.  

Our results highlight two primary physical mechanisms governing real-space magnetic exchange in the time-reversal breaking Weyl semimetal model. First, the spatial profile, magnitude, and anisotropic structure of the RKKY exchange tensor  remain remarkably unchanged across the Weyl node-annihilation transition. This demonstrates that short- and intermediate-range magnetic interactions are mediated by the global spin-momentum locking embedded across the full Brillouin zone, rather than being strictly dictated by the low-energy topological nodes. Second, the antisymmetric Dzyaloshinskii–Moriya (DM) component  exhibits a strong Fermi-level activation effect. At charge neutrality this DM component is heavily suppressed due to a parity cancellation of its integrand over the fully occupied valence band. Shifting the Fermi level past the Van Hove scale  breaks this domain integration symmetry, causing the DM component  to grow rapidly and compete directly in magnitude with the diagonal Heisenberg terms.

In addition, a key contribution of this work is the derivation of a generic, closed-form matrix formulation for the RKKY exchange tensor for generic two-band systems governed by Hamiltonians of the form $H(\mathbf{k}) = \mathbf{h}(\mathbf{k}) \cdot \boldsymbol{\sigma}$ in the spin basis. By expanding the real-space retarded Green’s function into its scalar identity and Pauli vector components,  we explicitly evaluate the full spin traces analytically. This reduces the complete $3 \times 3$ magnetic interaction tensor—encompassing isotropic Heisenberg, anisotropic Ising, Dzyaloshinskii–Moriya, and pseudo-dipolar couplings—into simple combinations of fundamental energy integrals. While Green’s function techniques are often applied to specific low-energy models, this unified algebraic reduction is exact across the entire Brillouin zone and holds for arbitrary two-band models, providing a ready-to-use framework for calculating real-space indirect RKKY exchange integrals.



The manuscript is organized as follows: Sec. \ref{sec:RKKYExpression} presents the derivation of the generic matrix formulation of the RKKY exchange tensor for arbitrary two-band models, while In Sec. \ref{sec:ModelMethods} introduces the  specific  two-band lattice model hosting WSMs, discusses its symmetry properties, and presents the corresponding real-space Green's functions and RKKY matrix elements. We present our results in Sec. \ref{sec:Results}: the numerical calculations of the RKKY exchange couplings are discussed, detailing their behavior across the node-annihilation threshold and under Fermi-level tuning. Finally, in Sec. \ref{sec:conclusion} we present our conclusion remarks. Some technical developments details and discussion are left to the Appendices.

\section{RKKY interaction expression for generic two-band models. }
\label{sec:RKKYExpression}

We begin by considering that the Weyl itinerant electrons interact with magnetic impurities ${\bf S}_{i}$ located at positions ${\bf{R}}_{i}$ via an $s$-$d$ exchange interaction $H_{I}=J_i{\bf S}_{i}\cdot\boldsymbol{\sigma}_{i}\delta({\bf r}-{\bf R}_{i})$, where $J_i$ is the strength of the $s$-$d$ exchange interaction, and $\boldsymbol{\sigma}=(\sigma_{x},\sigma_{y},\sigma_{z})$ is the vector of Pauli matrices.

As such, there will be an {\rm indirect} interaction between two magnetic impurities mediated by these itinerant electrons. This is usually referred to as the Ruderman-Kittel-Kasuya-Yosida interaction or the ``RKKY interaction''\cite{PhysRev.96.99,10.1143/PTP.16.45,PhysRev.106.893}.
At zero temperature, the RKKY Hamiltonian describing the effective interaction between two impurities can be computed by second-order perturbation theory as~\cite{mattis2006theory,mahan2012many}
\begin{equation}\label{eq:rkky}
    H_{\rm RKKY}=-\dfrac{J^{2}}{\pi}\mbox{Im}\int_{-\infty}^{\epsilon_{F}} d\epsilon \; \mbox{Tr} \left[ \left({\bf S_{1}\cdot \bm{\sigma}}\right)\mathbb{G}^{0}\left({\bf R},\epsilon^{+}\right)\left({\bf S_{2}\cdot \bm{\sigma}}\right)\mathbb{G}^{0}\left({\bf -R},\epsilon^{+}\right) \right] \; ,
\end{equation}
where we have defined $J \equiv J_1 = J_2$, and ${\bf R} \equiv {\bf R}_{2}-{\bf R}_{1}$, $\epsilon_{F}$ is the Fermi energy, $\epsilon^{+}=\epsilon+i\delta~(\delta\rightarrow 0^{+})$, and $\mathbb{G}^{0}(\pm {\bf R}, \epsilon^{+})$ is the (non-interacting) host matrix Green's function in real space:
\begin{equation}
 \mathbb{G}(\pm {\bf R},\epsilon^{+})=\int\frac{d^{3}\veck}{(2 \pi)^3}e^{\pm i \veck\cdot {\bf R}}[\epsilon^{+} \sigma_0 -H_0(\veck)]^{-1} \; ,
\label{eq:defGFH0}
\end{equation}
where  $\sigma_0 \equiv \mathbbm{1}_{2 \times 2}$ is the unity $2\times2$ matrix, and $H_0(\veck)$ is the (non-interacting) 2-band Hamiltonian describing the itinerant electrons with momentum $\veck$. 

Equivalently, we can write $H_{\rm RKKY}$ in the compact form
\begin{equation}\label{eq:rkky2}
    H_{\rm RKKY}=\left[{\bf S_{1}}\right]^{T} {\bf J}_{\rm RKKY}({\bf R}) \; {\bf S_{2}}  ,
\end{equation}
where ${\bf J}_{\rm RKKY}({\bf R})$ is a 3 $\times$ 3 matrix whose elements $J_{ab}({\bf R})$ $(a,b=X,Y,Z)$ are given by
\begin{equation}\label{eq:Jrkky}
    J_{ab}({\bf R})=-\dfrac{J^{2}}{\pi}\mbox{Im}\int_{-\infty}^{\epsilon_{F}} d\epsilon \; \mbox{Tr} \left[ \left(\bm{\sigma}\right)_a \mathbb{G}\left({\bf R},\epsilon^{+}\right)\left(\bm{\sigma}\right)_b\mathbb{G}\left({\bf -R},\epsilon^{+}\right) \right]  ,
\end{equation}


Our first goal is to write a generic expression for $H_{\rm RKKY}$ defined in Eq.~\eqref{eq:rkky} and \eqref{eq:rkky2} when the itinerant electrons are described by generic spinful two-band Hamiltonians of the form:
\begin{equation}
H_0(\veck)=h_{x}(\veck) \sigma_{x}+h_{y}(\veck) \sigma_{y}+h_{z}(\veck) \sigma_{z} \equiv \bm{h}(\veck) \cdot \bm{\sigma}.
\label{eq:generic_H0}
\end{equation}
where 
$\sigma_{i=x,y,z}$ are Pauli matrices in spin space.

Using the identity 
\begin{equation}
[\epsilon^{+} \sigma_0 -H_0(\veck)]^{-1} = \frac{\left((\epsilon^{+} )\sigma_0 + \bm{h}(\veck) \cdot \bm{\sigma}\right)}{\left( (\epsilon^{+} )^2 - |\bm{h}(\veck)|^2 \right)} \; ,
\label{eq:Identity}
\end{equation}
the matrix Green's function defined in Eq.~\eqref{eq:defGFH0} can be written as follows:  
\begin{equation}
\mathbb{G}(\pm {\bf R},\epsilon^{+})=G_{0}(\pm {\bf R},\epsilon^{+})\s_{0}+G_{x}(\pm {\bf R},\epsilon^{+})\s_{x}+G_{y}(\pm {\bf R},\epsilon^{+})\s_{y}+G_{z}(\pm {\bf R},\epsilon^{+})\s_{z} \; ,
\label{eq:GFH0}
\end{equation}
where the terms $G^{0}_{i}(\pm {\bf R},\epsilon^{+})$ will be given by
\begin{eqnarray}
 G_{0}(\pm{\bf R},\epsilon^{+}) &=\int_{\rm BZ}\frac{d^{3}{\bf k}}{(2 \pi)^3}e^{\pm i{\bf k}\cdot{\bf R}}\frac{\epsilon^{+}}{\left(\epsilon^{+} \right){}^{2}-|\bm{h}(\veck)|^{2}},  \label{eq:greens_2band_sigma0} \\
 G_{x}(\pm{\bf R},\epsilon^{+}) &=\int_{\rm BZ}\frac{d^{3}{\bf k}}{(2 \pi)^3}e^{\pm i{\bf k}\cdot{\bf R}}\frac{h_{x}(\veck)}{\left(\epsilon^{+}\right){}^{2}-|\bm{h}(\veck)|^{2}}, \label{eq:greens_2band_sigmax}\\
  G_{y}(\pm{\bf R},\epsilon^{+}) &=\int_{\rm BZ}\frac{d^{3}{\bf k}}{(2 \pi)^3}e^{\pm i{\bf k}\cdot{\bf R}}\frac{h_{y}(\veck)}{\left(\epsilon^{+}\right){}^{2}-|\bm{h}(\veck)|^{2}},\label{eq:greens_2band_sigmay} \\
  G_{z}(\pm{\bf R},\epsilon^{+}) &=\int_{\rm BZ}\frac{d^{3}{\bf k}}{(2 \pi)^3}e^{\pm i{\bf k}\cdot{\bf R}}\frac{h_{z}(\veck)}{\left(\epsilon^{+} \right){}^{2}-|\bm{h}(\veck)|^{2}} \; . \label{eq:greens_2band_sigmaz}
\end{eqnarray}

Finally, the matrix elements $J_{ab}({\bf R})$ defined in Eq.~\eqref{eq:Jrkky} can be written in terms of combinations of the real and imaginary parts of integrals involving products of the $G^{0}_{i}(\pm {\bf R},\epsilon^{+})$, defined as 
\begin{align}
    j_{ij}({\bf R})&=-\dfrac{J^{2}}{\pi}\mbox{Im}\int_{-\infty}^{\epsilon_{F}}  G_{i}(-{\bf R},\epsilon^{+}) G_{j}(+{\bf R},\epsilon^{+}) \; d\epsilon \; , \label{eq:jij} \\
    j^{\rm Re}_{ij}({\bf R})&=-\dfrac{J^{2}}{\pi}\mbox{Re}\int_{-\infty}^{\epsilon_{F}}  G_{i}(-{\bf R},\epsilon^{+}) G_{j}(+{\bf R},\epsilon^{+}) \; d\epsilon \; ,  \label{eq:Rejij}
\end{align}

We notice that the connection between Eqs.~\eqref{eq:jij}-\eqref{eq:Rejij} and Eq.~\eqref{eq:Jrkky} implies that the spatial symmetries (or lack thereof)  of the Hamiltonian will be reflected in symmetries/anti-symmetries in the  RKKY interaction matrix ${\bf J}_{\rm RKKY}({\bf R})$. A specific example will be given in Section \ref{sec:TwoBandModel}.

\section{RKKY interaction for a TR-breaking Weyl semimetal}
\label{sec:ModelMethods}

\subsection{Two-band tight-binding model}
\label{sec:TwoBandModel}

As a starting point for our analysis, we will consider the spinful two-band tight-binding model in a cubic lattice  proposed in Refs.~\cite{PhysRevB.86.214514,PhysRevB.97.125419,yang}. In the notation of Eq.~\eqref{eq:generic_H0}, the model reads
\begin{equation}\label{h_k}
H^{\rm W}_0(\veck) = \bm{h}(\veck) \cdot \bm{\sigma} = t \sin(k_{x})\sigma_{x}+ t \sin(k_{y})\sigma_{y}+h^{m,k_0}_z(\veck)\sigma_{z} \; ,
\end{equation}
where $h^{m,k_0}(\veck)=t\left[\cos(k_{z})-\cos(k_{0})\right]+m[2-\cos(k_{y})-\cos(k_{x})]$ vanishes at $\veck^{W}_{\pm}=(0,0,\pm k_0)$.  

This model breaks time-reversal symmetry (TRS), i.e., $\mathcal{T} H^{\rm W}_0(-\veck) \mathcal{T}^{-1} \neq H^{\rm W}_0(\veck)$ where $\mathcal{T}=i\sigma_y \hat{K}$ is the time-reversal operator. More concretely, the TRS breaking arises due to $\mathcal{T} h^{m,k_0}(\veck) \sigma_z \mathcal{T}^{-1} = - h^{m,k_0}(\veck) \sigma_z$ for any value of $k_0$ and $m$. 

The energies of the two bands given by $E_{\pm}\left({\bf k}\right)=\pm |\bm{h}(\veck)|$ with  
\begin{equation}
\label{eq:EkTwoBand}
 |\bm{h}(\veck)|= t \sqrt{\sin^{2}(k_{x})+\sin^{2}(k_{y})+[(\cos(k_{z})-\cos(k_0))+(m/t)(2-\cos(k_{x})-\cos(k_{y}))]^{2}}.
\end{equation}

\begin{figure}[h!]
\centering
\includegraphics[width=0.45\columnwidth]{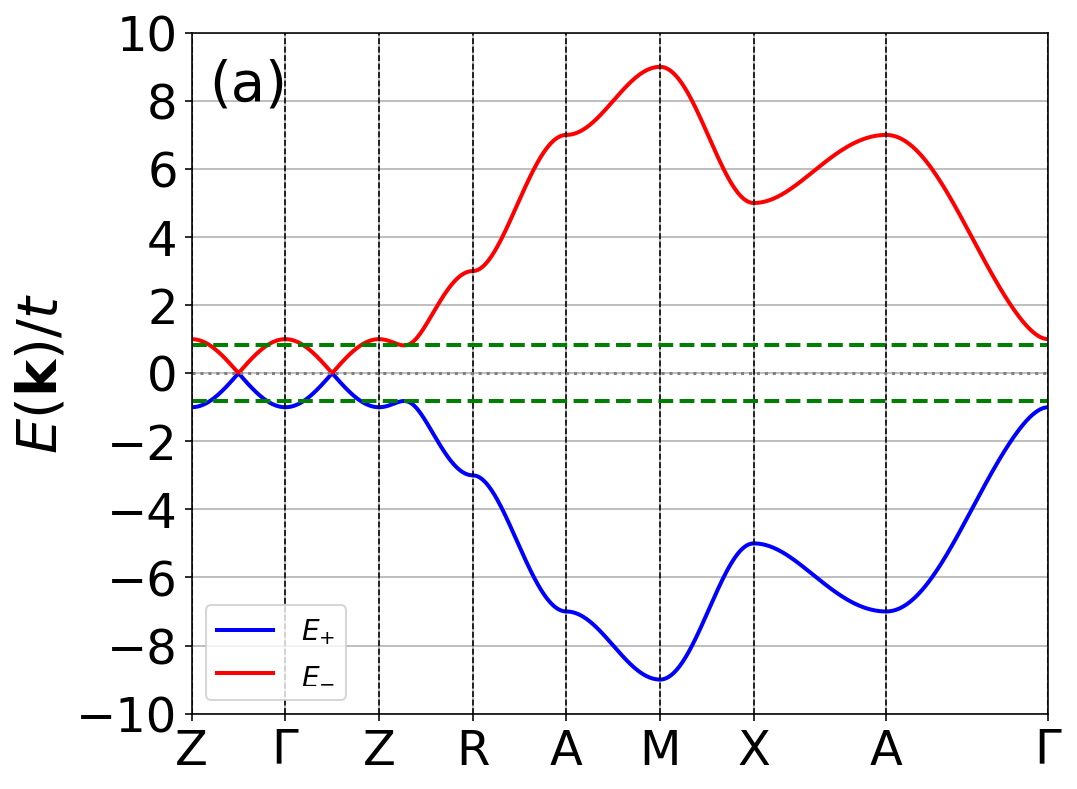}
\includegraphics[width=0.45\columnwidth]{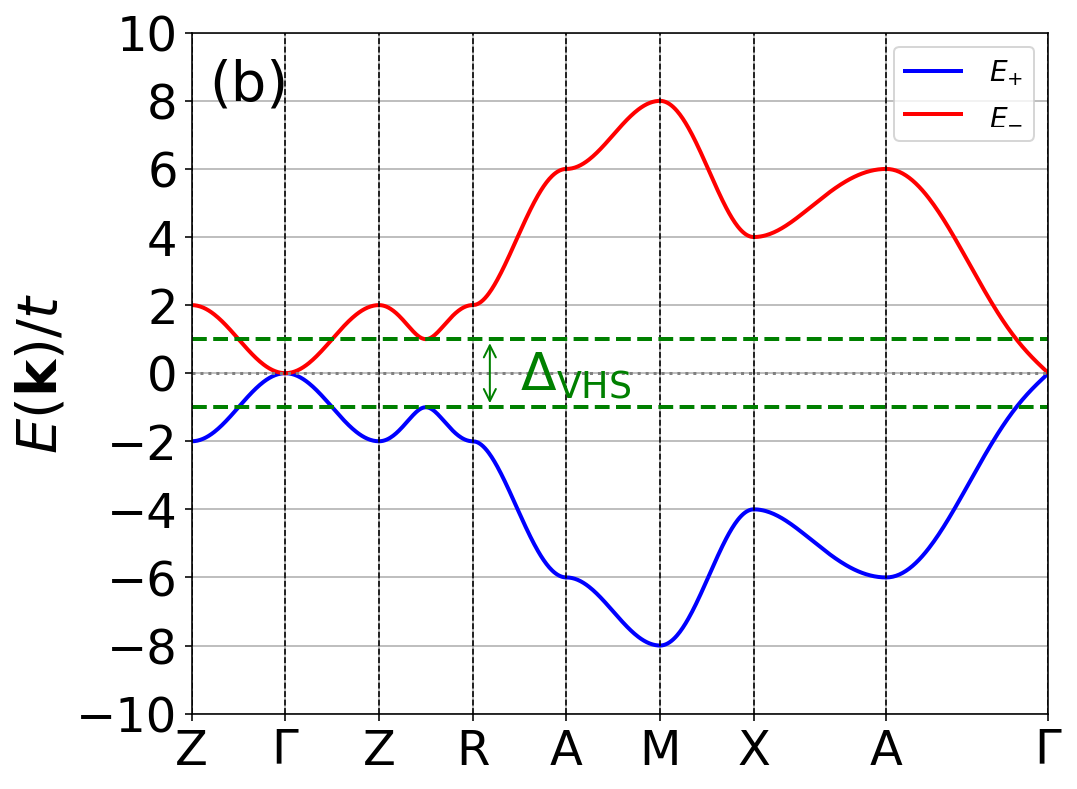}
\caption{Band structure along the high-symmetry points of the Brillouin zone for the two-band tight-binding model with $m=2$ and (a) $k_0=\pi/2$, (b) $k_0=0$. Dashed green lines mark the energies of the Van-Hove singularities, separated by an energy interval $\Delta_{\rm VHS}$. Notice that the well-defined Weyl nodes at $(0,0,\pm \pi/2)$ in (a) have merged into a quadratic band-touching at the $\Gamma$ point in (b).}
\label{fig:BandStructure_mEQ2}
\end{figure}

The bands are symmetric with respect to inversion about the $\Gamma$ point [$E_{\pm}\left({\bf k}\right)=E_{\pm}\left(-{\bf k}\right)$] and along the $x=y$ plane [$E_{\pm}\left(k_x,k_y,k_z\right)=E_{\pm}\left(k_y,k_x,k_z\right)$]. For a non-zero Wilson mass term ($m>0$), the  model sustains two Weyl nodes with opposite chiralities at the momentum points $\veck^{W}_{\pm}=(0,0,\pm k_0)$. This is as depicted in Figure \ref{fig:BandStructure_mEQ2}(a), showing the energy bands  $E_{\pm}(\veck)$ for $m=2$ and $k_0=\pi/2$ along the high-symmetry points (HSPs) in the (cubic) Brillouin zone. For $k_0=0$ (Fig.~\ref{fig:BandStructure_mEQ2}(b)), the two Weyl nodes merge, rendering a quadratic band touching at the $\Gamma$ point.

Notice that for large enough values of $m$ (such as $m=2$ used in Fig.~\ref{fig:BandStructure_mEQ2}), the bands touch only at the Weyl points $\veck^{W}_{\pm}=(0,0,\pm k_0)$ and the system remains gapped at all other BZ points, as shown by the dashed green lines in  Fig.~\ref{fig:BandStructure_mEQ2}). The size of this ``gap" can be estimated as $\Delta_{\rm VHS} \equiv E_{+}(\veck_{\rm VHS})-E_{-}(\veck_{\rm VHS})$, where $\veck_{\rm VHS}=(\bar{k},0,\pi)$ and $\bar{k}=\mbox{arccos}\left[ 2(1-\cos{k_0})/3\right]$ marks a Van-Hove singularity $\left. \nabla^2 E(\veck) \right|_{\veck=\veck_{\rm VHS}}=0$.


\subsection{RKKY interaction matrix}
\label{sec:RKKYmatrix}

The real-space Green's functions (RSGFs) for this model are given by
\begin{align}
 G_{0}(\pm{\bf R},\epsilon^{+})&=\int_{\rm BZ}\frac{d^{3}{\bf k}}{(2 \pi)^3}e^{\pm i{\bf k}\cdot{\bf R}}\frac{\epsilon^{+}}{\left(\epsilon^{+}\right){}^{2}-|\bm{h}(\veck)|^{2}},  \label{eq:greens_2bandWeyl_sigma0} \\
 G_{x}(\pm{\bf R},\epsilon^{+})&=\int_{\rm BZ}\frac{d^{3}{\bf k}}{(2 \pi)^3}e^{\pm i{\bf k}\cdot{\bf R}}\frac{t\sin(k_{x})}{\left(\epsilon^{+}\right){}^{2}-|\bm{h}(\veck)|^{2}}, \label{eq:greens_2bandWeyl_sigmax}\\
  G_{y}(\pm{\bf R},\epsilon^{+})&=\int_{\rm BZ}\frac{d^{3}{\bf k}}{(2 \pi)^3}e^{\pm i{\bf k}\cdot{\bf R}}\frac{t\sin(k_{y})}{\left(\epsilon^{+}\right){}^{2}-|\bm{h}(\veck)|^{2}},\label{eq:greens_2bandWeyl_sigmay} \\
  G_{z}(\pm{\bf R},\epsilon^{+})&=\int_{\rm BZ}\frac{d^{3}{\bf k}}{(2 \pi)^3}e^{\pm i{\bf k}\cdot{\bf R}}\frac{t[\cos(k_{z})-\cos(k_{0})]+m[2-\cos(k_{y})-\cos(k_{x})]}{\left(\epsilon^{+}\right){}^{2}-|\bm{h}(\veck)|^{2}}. \nonumber \\ \label{eq:greens_2bandWeyl_sigmaz}
\end{align}
with $|\bm{h}(\veck)|$ given by Eq. \eqref{eq:EkTwoBand}. The integrals are performed within the cubic Brillouin zone of the model.

For simplicity, we will consider impurities located at positions along the  diagonal of the cubic primitive cell such that ${\bf R}=(R,R,R)$ and $|{\bf R}|=\sqrt{3} R$. In this case, due to the $x-y$ band symmetry, the RKKY interaction matrix for the model will be of the form:
\begin{equation}
{\bf J}_{\rm RKKY}(R) = 
\left(  
\begin{array}{rcl}
J_{XX}(R) &  J_{XY}(R) & J_{XZ}(R) \\
J_{XY}(R)&  J_{XX}(R) & -J_{XZ}(R) \\
-J_{XZ}(R) &  J_{XZ}(R) & J_{ZZ}(R)  
\end{array}
\right)
\label{eq:RKKYmatrix}
\end{equation}
with the matrix elements given in terms of the integrals  defined in Eqs. \eqref{eq:jij} and \eqref{eq:Rejij} as
\begin{align}
J_{XX}({\bf R}) &= j_{00}({\bf R}) - j_{zz}({\bf R}) \label{eq:JXX} \\
J_{ZZ}({\bf R}) &= j_{00}({\bf R}) -2j_{xy}({\bf R}) + j_{zz}({\bf R})  \label{eq:JZZ}\\
J_{XY}({\bf R}) &= 2 j_{xy}({\bf R})  \label{eq:JXY}\\
J_{XZ}({\bf R}) &= -2 j^{\rm Re}_{0x}({\bf R})  \label{eq:JXZ} 
\end{align}

Finally, the RKKY spin interaction Hamiltonian [Eq.~\eqref{eq:rkky2}] can then be expressed as a sum of Heisenberg XXZ, Dzyaloshinskii–Moriya and frustration terms in the form $H_{\rm RKKY}=H_{\rm XXZ}+H_{\rm DM}+H_{\rm XY}$ with

\begin{align}
H_{\rm XXZ} &= J_{XX}(R)\left( S_{1x}S_{2x} + S_{1y}S_{2y}  \right) +  J_{ZZ}(R)S_{1z}S_{2z} \label{eq:HXXZ}\\
H_{\rm DM} &= -J_{XZ}(R)\left[ {\bf S_{1}} \times {\bf S_{2}} \right] \cdot \left( {\bf i}+{\bf j} \right) \label{eq:HDM}\\
H_{\rm XY} &= J_{XY}(R) \left( S_{1x}S_{2y} + S_{1y}S_{2x}  \right) \label{eq:HXY}
\end{align}


\section{Numerical Results}
\label{sec:Results}

Our aim is to numerically calculate the RKKY couplings $J_{ab}(R)$  of the spin interaction Hamiltonian [Eq.~\eqref{eq:rkky2}] given by  Eqs.~\eqref{eq:HXXZ}--\eqref{eq:HXY}. 

The procedure is as follows: we start by numerically calculating the RSGFs  given by Eqs.~\eqref{eq:greens_2bandWeyl_sigma0}--\eqref{eq:greens_2bandWeyl_sigmaz} as a function of $R$ (which is proportional to the impurity separation $|{\bf R}|=\sqrt{3} R$) and energy $\epsilon$. In all cases, a cubic grid of $(N_k)^3$ points with $N_k=400$, an energy spacing  $\Delta _\epsilon = 0.005 t $ and a fixed broadening $\delta=0.05 t$ are used to obtain the RSGFs. Additional details of the numerical procedure and convergence (including the study of the benchmark given by the usual analytical expression for the RKKY interaction in the metallic case) are given in \ref{sec:TBbenchmark}. 

Then, for each value of $R$, we perform the energy integrals  defined in Eqs. \eqref{eq:jij} and \eqref{eq:Rejij} in order to obtain $j_{ij}(R)$ and $j^{\rm Re}_{ij}(R)$. Finally, we calculate the RKKY matrix elements $J_{ab}(R)$ using Eqs.~\eqref{eq:JXX}--\eqref{eq:JXZ}. By combining Eqs. \eqref{eq:jij} and \eqref{eq:Rejij} and Eqs.~\eqref{eq:JXX}--\eqref{eq:JXZ}, it is also useful to express $J_{ab}(R)$ in a compact form:
\begin{equation}
    J_{ab}(R) =-\dfrac{J^{2}}{\pi}\int_{-\infty}^{\epsilon_{F}} F_{ab}(R,\epsilon^{+})  \; d\epsilon \; , \label{eq:JabIntegral}
\end{equation}

\begin{figure}[h!]
\centering
\includegraphics[width=0.47\columnwidth]{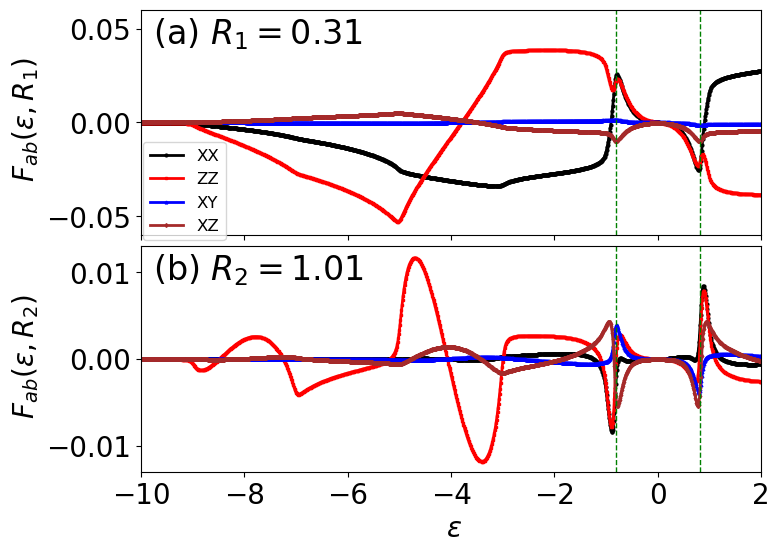}
\includegraphics[width=0.45\columnwidth]{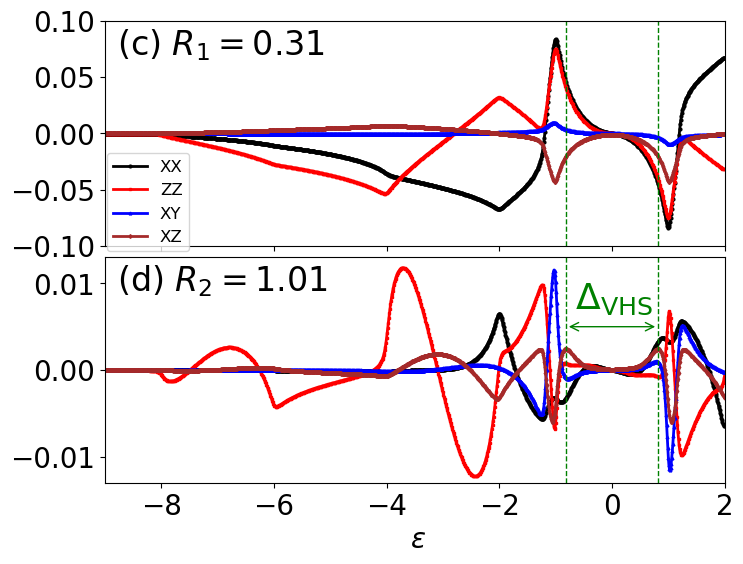}
\caption{Integrand of Eqs. \eqref{eq:JabIntegral} for $m=2$ and (a,b) $k_0=\pi/2$ (Weyl semimetal) (c,d) $k_0=0$ (merged Weyl cones) and for $R=0.31$ (a,c) and $R=1.01$ (b,d). The vertical dashed lines indicate the energies of the Van-Hove singularities, separated by  $\Delta_{\rm VHS}$. }
\label{fig:Gi2_k0EQPio2_0}
\end{figure}

The integrands $F_{ab}(R,\epsilon^{+})$ are plotted in Fig.~\ref{fig:Gi2_k0EQPio2_0} as a function of $\epsilon$ and two values of $R$. Notice that, for a given $R$, $F_{XX}(R,\epsilon^{+})$, $F_{ZZ}(R,\epsilon^{+})$ and $F_{XY}(R,\epsilon^{+})$ are odd functions of $\epsilon$ while $F_{XZ}(R,\epsilon^{+})$ (related to the Dzyaloshinskii–Moriya term) is an even function  of $\epsilon$. This can be traced back to  Eqs.~\eqref{eq:Rejij} and \eqref{eq:JXZ}  and the fact that the $XZ$ component is related to the real part of the product of RSGFs.

Notice also that the $F_{ab}(R,\epsilon^{+})$ displays a strong dependence for energies outside the VHS gap ($|\epsilon|>\Delta_{VHS}$), while tends to flatten out inside the energy interval within the VHS gap  ($|\epsilon|<\Delta_{VHS}$). Moreover, $F_{ab}(R,\epsilon^{+})$ is strongly suppressed for energies larger than the band-width of the model ($|\epsilon|\gtrsim 9t$ for the parameters of Fig.~\ref{fig:Gi2_k0EQPio2_0}). As such, the bandwidth of the model provides a natural energy cutoff above which the RSGFs in  are suppressed due to the quadratic denominator in Eqs.~\eqref{eq:greens_2bandWeyl_sigma0}--\eqref{eq:greens_2bandWeyl_sigmaz}. 

We notice in passing that the presence of a natural energy cutoff in band models provides a sharp contrast to RKKY calculations for  Weyl Hamiltonians which consider only the low-energy linear dispersion of the cones, disregarding band-edge effects \cite{Chang:Phys.Rev.B:241103:2015,hosseini}. In such ``two-cone approximation", either the integration in energy is carried out assuming that the linear dispersion extends down to $\epsilon \rightarrow - \infty$ or an ad-hoc energy cutoff function is added to the energy integral \cite{hosseini,Saremi_PhysRevB.76.184430_2007}. 

As detailed in Ref. \cite{Saremi_PhysRevB.76.184430_2007}, such cutoff procedure needs to be done with care as to avoid artifacts in the results. Here, the natural cutoff provided by the band structure avoids this issue, provided that the energy integrals  over the filled bands (Eqs. \eqref{eq:JabIntegral}) are performed with a cutoff at or below the band-edge. A more detailed discussion on the dependence of the results with the energy cutoff is provided in \ref{sec:cutoff}.

\subsection{RKKY interactions}
\label{sec:GFs}

In this section, we show results for the RKKY interaction matrix elements (Eq.~\eqref{eq:RKKYmatrix}) for the limiting cases of well-defined Weyl cones ($k_0=\pi/2$, see Fig.~\ref{fig:BandStructure_mEQ2}(a)) and merged cones, with a parabolic band-touching ($k_0=0$, Fig.~\ref{fig:BandStructure_mEQ2}(b)). 

\begin{figure}[h!]
\centering
\includegraphics[width=0.45\columnwidth]{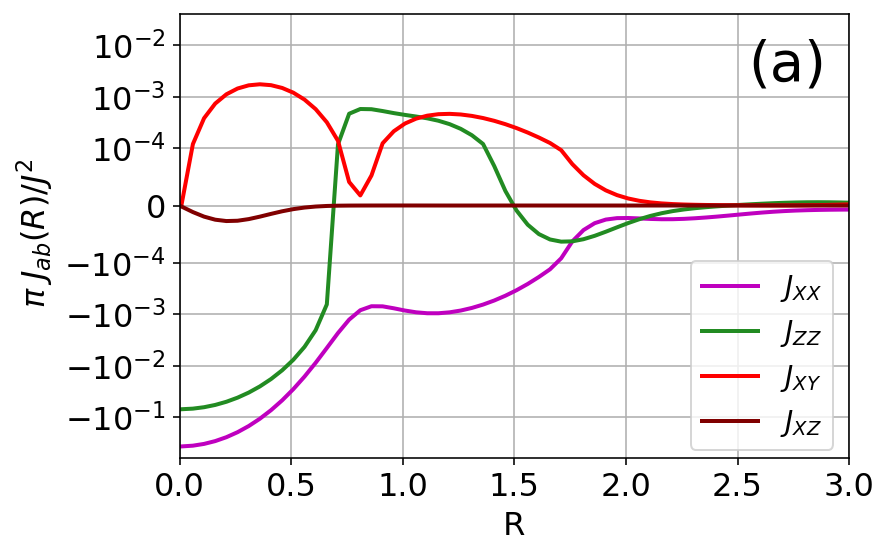}
\includegraphics[width=0.45\columnwidth]{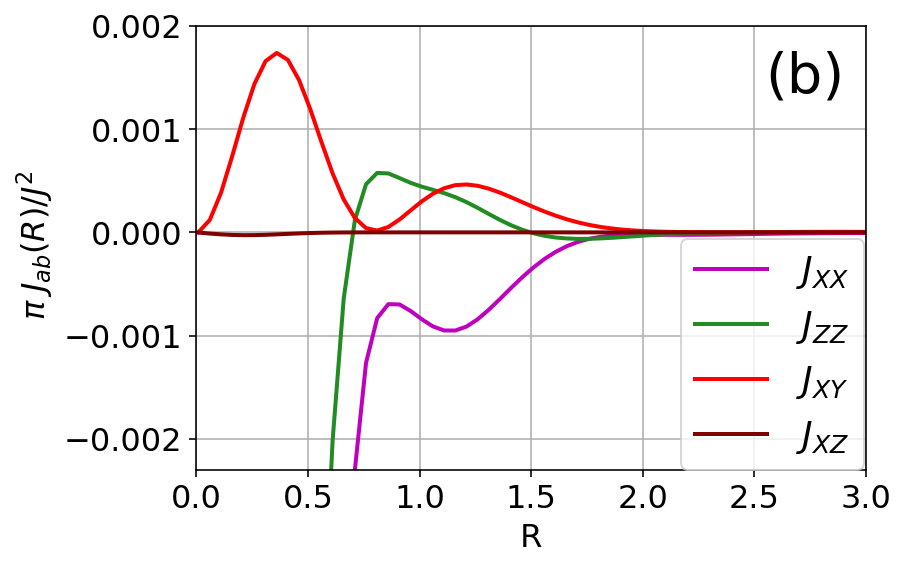}
\caption{RKKY matrix elements in (a) log-linear (b) linear-linear scale  for $\epsilon_{F}=0$, $m=2$ and $k_0=\pi/2$.}
\label{fig:RKKY_lolinear}
\end{figure}

We start our discussion with the case of $k_0=\pi/2$ and charge neutrality ($\epsilon_{F}=0$), such that the Fermi level is tuned exactly to the energy of the Weyl points  $(0,0,\pm \pi/2)$. These are shown in both log (Fig.~\ref{fig:RKKY_lolinear}(a)) and linear (Fig.~\ref{fig:RKKY_lolinear}(b)) scales.

As a general trend, the diagonal terms $J_{XX}(R)$ and $J_{ZZ}(R)$ dominate for small distances and favor a ferromagnetic alignment between the impurities, as in the regular metallic case (see also \ref{sec:TBbenchmark}). As the distance $R$ between the impurities increases, $J_{ZZ}(R)$ oscillates as a function of $R$, becoming antiferromagnetic for $R \approx 0.6$ (in units of the cubic lattice constant), while $J_{XX}(R)$ remains ferromagnetic for larger distances. By contrast, the off-diagonal frustation term $J_{XY}(R)$ remains antiferrogmanetic and becomes comparable to the diagonal ones for $ R \gtrsim 1$. The Dzyaloshinskii–Moriya term $J_{XZ}(R)$, by comparison, is negligible for these parameters.

\begin{figure}[h!]
\centering
\includegraphics[width=0.45\columnwidth]{J_vsR_k0EQPio2_mEQ2_EFEQ0_log.png}
\includegraphics[width=0.45\columnwidth]{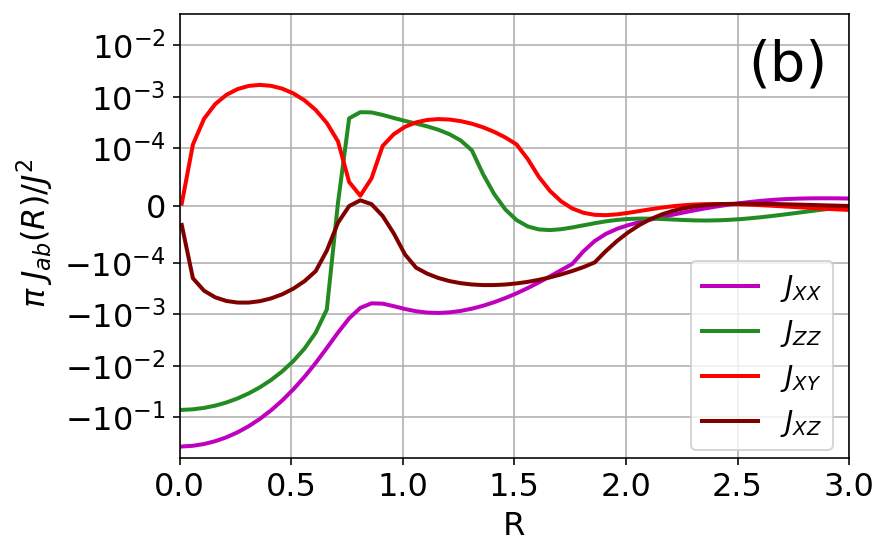}
\includegraphics[width=0.45\columnwidth]{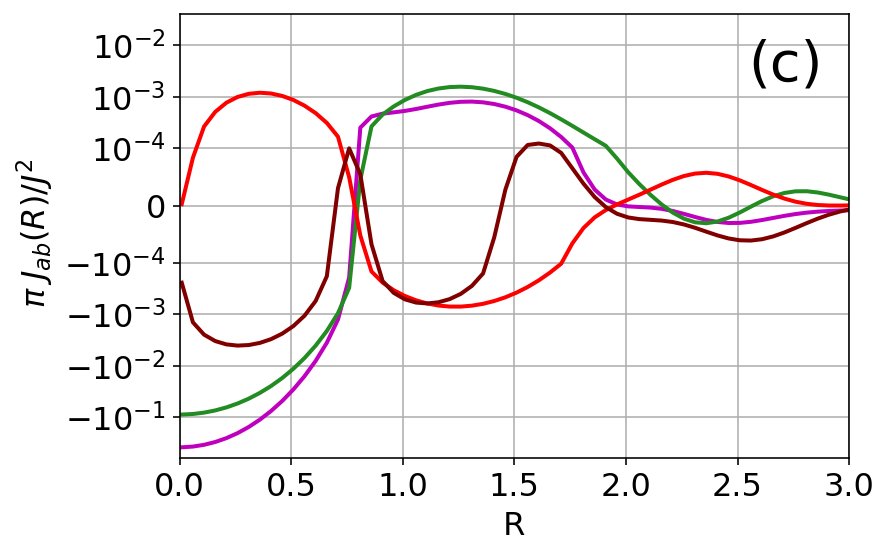}
\includegraphics[width=0.45\columnwidth]{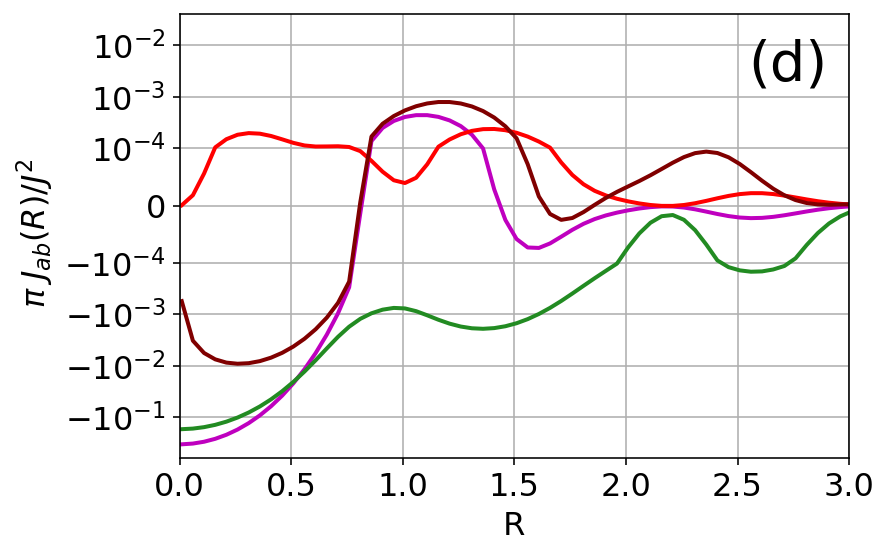}
\caption{RKKY matrix elements for $k_0=\pi/2$ and different values of the Fermi energy (a) $\epsilon_{F}=0$ (b) $\epsilon_{F}=0.5$ (c) $\epsilon_{F}=1.0$ (d) (c) $\epsilon_{F}=2.0$.}
\label{fig:RKKY_log_k0EQPio2}
\end{figure}

Figure \ref{fig:RKKY_log_k0EQPio2} explores the behavior of the RKKY interaction matrix the Fermi energy is varied away from charge neutrality. For Fermi energy values within the VHS gap ($|\epsilon_{F}|< \Delta_{\rm VHS}/2 \approx 1.0 t$, Figs.~\ref{fig:RKKY_log_k0EQPio2}(a-b)), the overall behavior is the same as in the charge neutral case ($\epsilon_{F}=0$). This can be understood by noticing that the contributions to the integrals given by Eq.~\eqref{eq:JabIntegral} is relatively small for this range of energies (see Fig.~\ref{fig:Gi2_k0EQPio2_0}(a,b)).

For larger values of $\epsilon_{F}$, however, the contribution of the higher bands to the integrals become important and significant changes occur in the RKKY matrix elements. For instance, $\epsilon_{F}=t$ (Fig.~\ref{fig:RKKY_log_k0EQPio2}(c)) the diagonal term $J_{XX}(R)$ oscillates with $R$ with a similar period and magnitude as $J_{ZZ}(R)$, rendering the XXZ as an Heisenberg-like term. Most notably, the off diagonal  Dzyaloshinskii–Moriya term $J_{XZ}(R)$ increases substantially, becoming of the same order as the diagonal ones for $R \gtrsim 1$. This trend continues for larger values of $\epsilon_{F}$ (Fig.~\ref{fig:RKKY_log_k0EQPio2}(d)), where the strong oscillatory behavior of the RKKY terms $J_{ab}(R)$ arises from the contribution of the high-energy bands.

The behavior of the Dzyaloshinskii-Moriya term $J_{XZ}(R)$ highlights a crucial symmetry-driven effect. As established in Eq.~\eqref{eq:JabIntegral} and Fig.~\ref{fig:Gi2_k0EQPio2_0}, the integrand $F_{XZ}(R, \epsilon^+)$ exhibits even parity with respect to $\epsilon$. At charge neutrality ($\epsilon_F = 0$), performing the energy integration over the filled symmetric valence band leads to a substantial cancellation of positive and negative spectral contributions, effectively suppressing $J_{XZ}$ at low energies. However, as the Fermi level is shifted away from neutrality—particularly when $\epsilon_F$ approaches or exceeds the Van Hove singularity scale $\Delta_{\text{VHS}}$ ($\epsilon_F \gtrsim 1.0t$)—this integration symmetry is broken. This Fermi-level tuning unleashes a strong antisymmetric exchange, allowing $J_{XZ}(R)$ to become comparable in magnitude to the diagonal Heisenberg components ($J_{XX}, J_{ZZ}$). This suggests that the DM interaction in these semimetals can be dynamically activated via electrostatic gating or chemical doping.

\begin{figure}[h!]
\centering
\includegraphics[width=0.45\columnwidth]{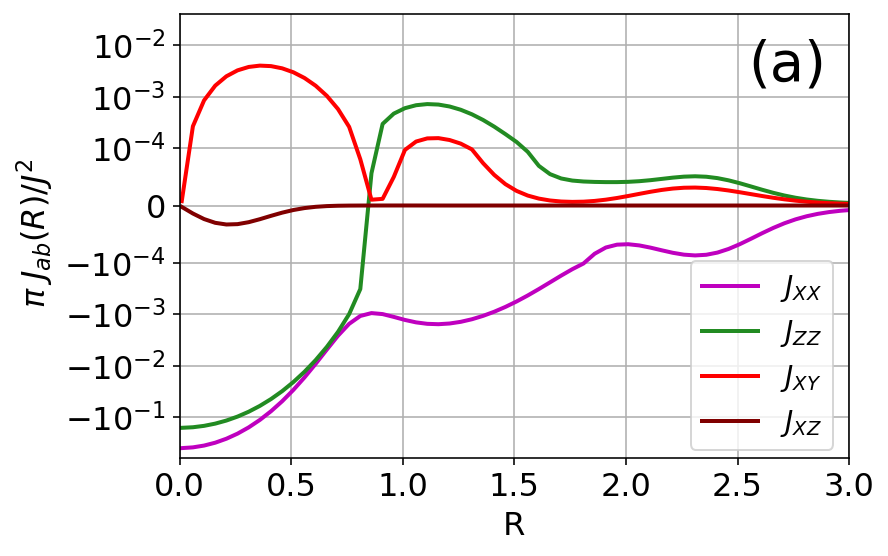}
\includegraphics[width=0.45\columnwidth]{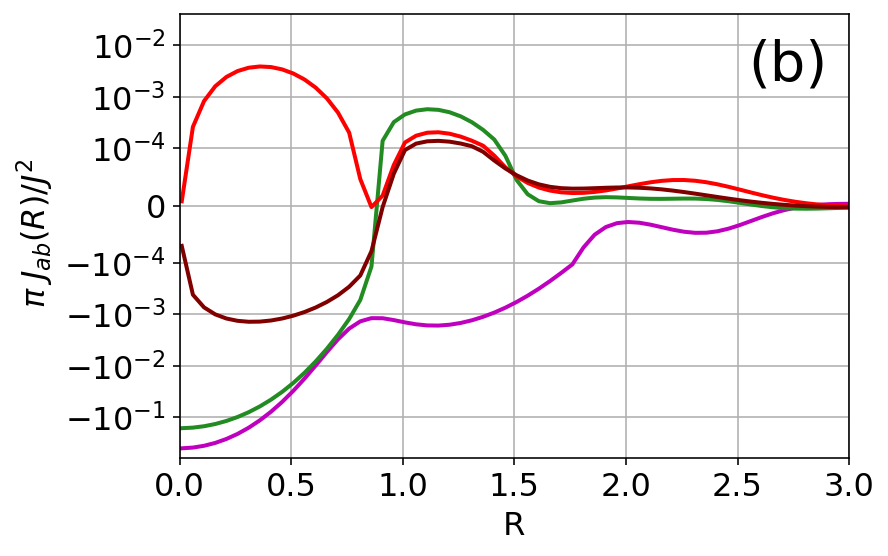}
\includegraphics[width=0.45\columnwidth]{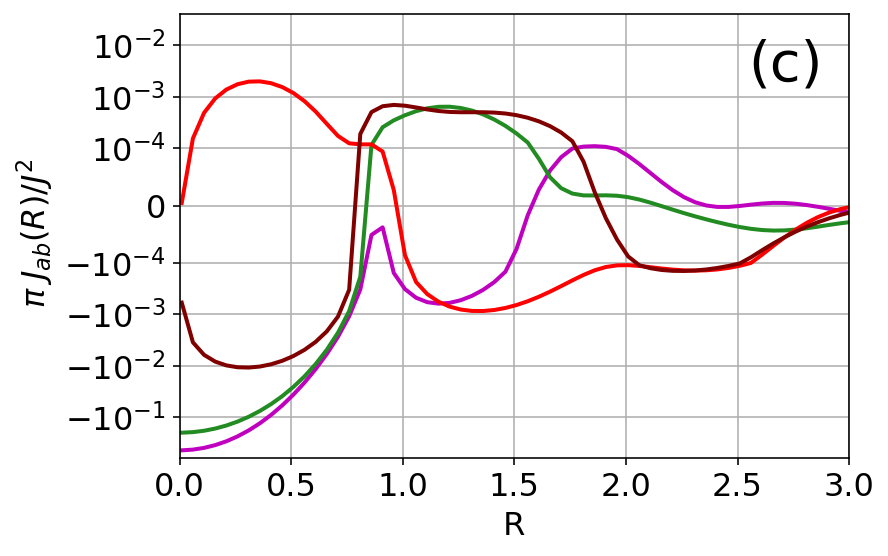}
\includegraphics[width=0.45\columnwidth]{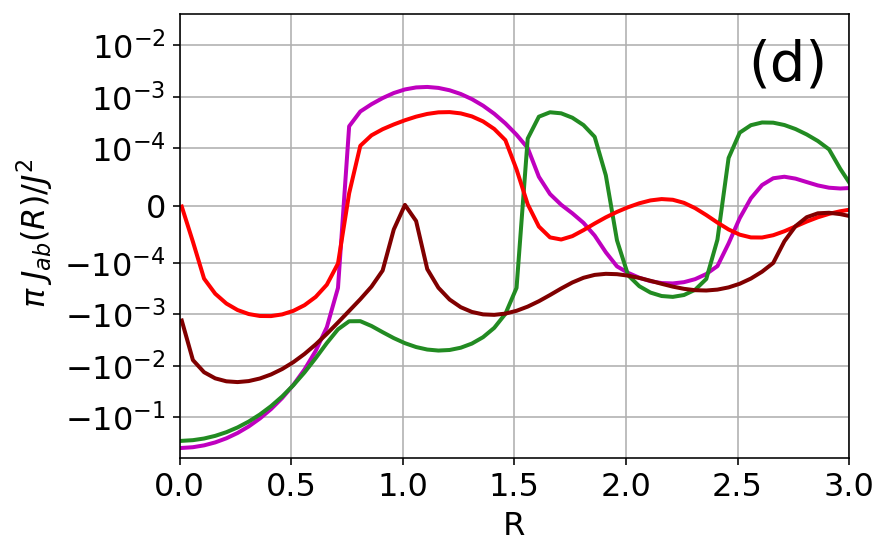}
\caption{RKKY matrix elements for $k_0=0$ and different values of the Fermi energy (a) $\epsilon_{F}=0$ (b) $\epsilon_{F}=0.5$ (c) $\epsilon_{F}=1.0$ (d) (c) $\epsilon_{F}=2.0$.}
\label{fig:RKKY_log_k0EQ0}
\end{figure}

More importantly, this overall behavior of the RKKY interaction is robust against the merging of the Weyl nodes attained by reducing $k_0$. This is shown in Fig.~\ref{fig:RKKY_log_k0EQ0}. The fact that the RKKY interaction is effectively unchanged by the presence or absence of Weyl points in the band structure is one of the main and intriguing results of this paper.

Remarkably, comparing the exchange couplings for separated Weyl nodes ($k_0 = \pi/2$, Fig.~\ref{fig:RKKY_log_k0EQPio2}) and merged nodes ($k_0 = 0$, Fig.~\ref{fig:RKKY_log_k0EQ0}), we observe that the overall magnitude and spatial profile of $J_{ab}(R)$ remain virtually unchanged. This robustness demonstrates that RKKY interactions at short-to-intermediate distances are primarily governed by the global spin-momentum locking of the full tight-binding band structure rather than the isolated low-energy topological nodes. While decreasing $k_0$ leads to a topological transition where the pair of chiral Weyl nodes annihilates into a quadratic band touching at $\Gamma$, the total occupied spectral weight mediating the $s$-$d$ exchange is conserved. Consequently, the spin-orbit texture encoded in $\mathbf{h}(\mathbf{k})\cdot \boldsymbol{\sigma}$ continues to dictate real-space spin precession, rendering the magnetic coupling robust across the node-annihilation threshold.

These findings demonstrate that RKKY interactions in this two-band model are governed by full-band electronic structure and energy-integration domain symmetry, rather than being strictly bound to isolated topological nodes. The persistence of the exchange profiles across $k_0 = \pi/2$ and $k_0 = 0$ confirms that spin-momentum locking survives the node-annihilation transition, preserving real-space spin textures across topological phase boundaries. At the same time, the transition of $J_{XZ}(R)$ from a symmetry-suppressed state at charge neutrality ($\epsilon_F = 0$) to a dominant exchange component at higher Fermi energies ($\epsilon_F \gtrsim \Delta_{\text{VHS}}$) highlights a practical mechanism for electric-field tuning of magnetic anisotropy. Together, these insights emphasize the necessity of full-band tight-binding approaches over low-energy linearized models when predicting real-space magnetic interactions in topological semimetals.

\section{Concluding remarks}\label{sec:conclusion}

In this work, we have developed a unified matrix framework to evaluate the full RKKY interaction tensor in generic two-band lattice systems. By analytically performing the spin traces, our approach reduces the complete real-space magnetic exchange tensor—encompassing isotropic Heisenberg, anisotropic Ising, Dzyaloshinskii–Moriya, and pseudo-dipolar couplings—to standard energy integrals over real-space Green's functions. Applying this methodology to a full tight-binding model bypasses the known artifacts of low-energy linearized continuum approximations, eliminating the need for artificial ultraviolet regulators and capturing the essential contributions of high-energy valence states across the entire Brillouin zone.

Applying this framework to a time-reversal-symmetry-breaking semimetal model, we demonstrated that the spatial profile, anisotropic structure, and overall magnitude of the exchange tensor remain remarkably persistent as the system transitions from a Weyl semimetal with well-separated chiral nodes to a quadratic band-touching semimetal via node annihilation. This topological robustness reveals that short- and intermediate-range magnetic exchange couplings are mediated primarily by the global spin-momentum locking embedded throughout the full electronic band structure, rather than being uniquely dictated by the local topology of isolated low-energy nodes.

Simultaneously, we uncovered a  mechanism controlling the strength of the symmetry-allowed Dzyaloshinskii–Moriya interaction in this system. At charge neutrality, the Dzyaloshinskii–Moriya component is heavily suppressed due to exact parity cancellations when integrating over the fully occupied, symmetric valence band. As the Fermi level is shifted away from neutrality toward or past the Van Hove singularity scale, this integration domain symmetry is broken, unmasking latent spectral contributions and allowing the Dzyaloshinskii–Moriya exchange to grow rapidly until it matches the diagonal Heisenberg terms in magnitude. This Fermi-level activation offers a practical knob for gate-tuning magnetic anisotropy in topological semimetals.

Moreover, our results provide an interesting perspective on the quantum geometric nature of indirect exchange in lattice models \cite{oh2026quantumgeometryrkkyflat}. Although local topological transitions are governed by the annihilation of isolated Berry curvature singularities, the real-space RKKY interaction is fundamentally mediated by wave-function overlaps summed across all occupied states, a feature encoded in the real part of the quantum geometric tensor—the quantum metric \cite{oh2026quantumgeometryrkkyflat}. As such, the fact that the real-space RKKY spin-exchange remains practically unaltered upon the convergence of the Wyl nodes indicates that the global, Brillouin-zone-integrated, quantum metric is essentially conserved across the node-annihilation threshold. This indicates that RKKY interactions reflect the global quantum metric of the full electronic spectrum rather than being strictly bound to the local low-energy Berry curvature flux.

While our framework provides a transparent full-band mapping of the RKKY exchange tensor for generic two-band models, there are valuable directions for future investigations. First, our calculations are evaluated at zero temperature and incorporating finite-temperature thermal broadening via Matsubara techniques would be useful to quantify the thermal damping of the spatial oscillations at larger impurity separations. Second, 
realistic topological semimetals often host complex electronic structures with coexisting trivial Fermi pockets or multiple pairs of nodes. In this sense, extending this algebraic scheme to multi-band tight-binding or \textit{ab initio} DFT-derived Hamiltonians will enable direct, material-specific predictions of real-space magnetic anisotropy in candidate compounds, providing precise exchange parameters for effective spin models and Kondo-lattice Hamiltonians.  

Collectively, these findings emphasize the necessity of full-band electronic structure considerations when modeling real-space magnetic interactions in topological matter. Quantifying the spatial decay and tensor anisotropy of the exchange interactions provides essential baseline input for describing heavy-fermion physics and correlated state transitions, such as those governing the competition between local Kondo screening and long-range magnetic order in candidate Weyl-Kondo semimetals.

\section*{Acknowledgments}

We acknowledge fruitful discussions with Thaís Trevisan and Eduardo Miranda, as well as financial support from Brazilian funding agencies CAPES, CNPq (grants No. 309789/2020-6, and 312622/2023-6), and FAPESP (grants No. 2021/04078-0 and 2022/15453-0).

\appendix

\section{Details of the numerical calculations}
\label{sec:appendix}

In this section, we provide additional details on the numerical calculations included in the paper. This includes the spacings used in the numerical integrations of  Eqs.~\eqref{eq:greens_2bandWeyl_sigma0}--\eqref{eq:greens_2bandWeyl_sigmaz} and \eqref{eq:JabIntegral}, as well as the convergence of $J_{ab}(R)$ defined in Eq.~\eqref{eq:JabIntegral} with the energy cutoff.

\subsection{Benchmark: 3D tight-binding model vs analytical results}
\label{sec:TBbenchmark}

In this section, we set the benchmark for numerical parameters (broadening, spacing in ${\bf k}$ space, spacing in energy, energy cutoff) and compare it with the well-known analytical expression for the RKKY interaction for a 3D electron gas with a quadratic dispersion \cite{coleman}:

\begin{equation}
J^{\rm exact}_{\rm RKKY}({\bf R}) = -J^2 \left( \frac{k_F^4}{2 \pi^3} \right) \left[ \frac{\sin(2k_F|{\bf R}|)-(2k_F |{\bf R}|)\cos(2k_F|{\bf R}|)}{(2k_F|{\bf R}|)^4} \right] \; ,
\label{eq:RKKYanalytic}
\end{equation}
where, in the units used, the dispersion is given (in units of $t$) by $E_{\rm free}(\veck) = t|\veck|^2 + E_0$   and $k_F=\sqrt{(\epsilon_{F}-E_0)/t}$ with $E_0$ being the energy at the bottom of the conduction band.

In order to validate the parameters for our tight-binding calculations, we compare $J^{\rm exact}_{\rm RKKY}({\bf R})$ given by Eq.~\eqref{eq:RKKYanalytic} with a numerical result obtained by evaluating Eq.~\eqref{eq:Jrkky} (or, equivalently, Eq.~\eqref{eq:jij}) for a single-band 3D tight-binding model:
\begin{equation}
 J^{\rm TB}_{\rm RKKY}({\bf R}) =-\dfrac{J^{2}}{\pi} \int_{E_{\rm cutoff}}^{\epsilon_{F}} \mbox{Im } \left[ G_{\rm TB}(-{\bf R},\epsilon^{+}) G_{\rm TB}(+{\bf R},\epsilon^{+}) \right] \; d\epsilon
\label{eq:RKKY_3DTB}
\end{equation}
where the RSGF is analogous to that given by Eq.~\eqref{eq:defGFH0}:
\begin{equation}
 G_{\rm TB}(\pm{\bf R},\epsilon^{+}) =\int_{\rm BZ} \frac{d^{3}{\bf k}}{(2 \pi)^3}\frac{e^{\pm i{\bf k}\cdot{\bf R}}}{\left(\epsilon+ i\delta -E_{\rm TB}(\veck) \right)} \;,  
 \label{eq:GF_TB} 
\end{equation}
with 
\begin{equation}
 E_{\rm TB}(\veck) = -2t \left[ \cos{k_x} + \cos{k_y} + \cos{k_z}\right]  \; .  
 \label{eq:ETB} 
\end{equation}

\begin{figure}[h!]
\centering
\includegraphics[width=0.45\columnwidth]{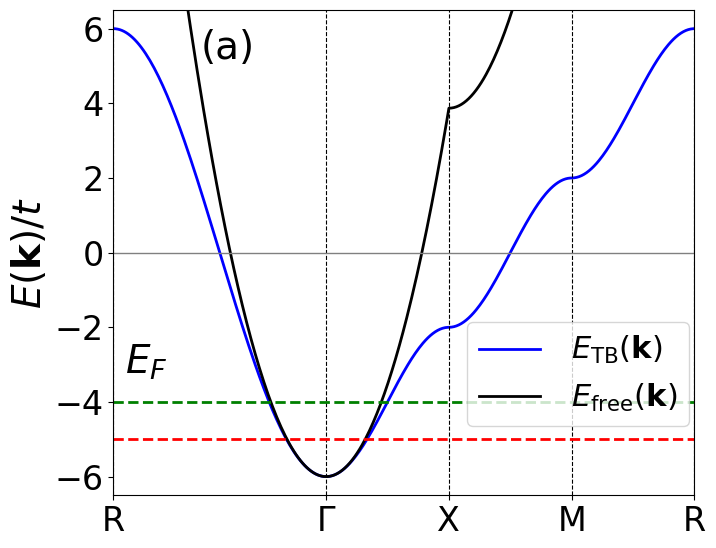}
\includegraphics[width=0.45\columnwidth]{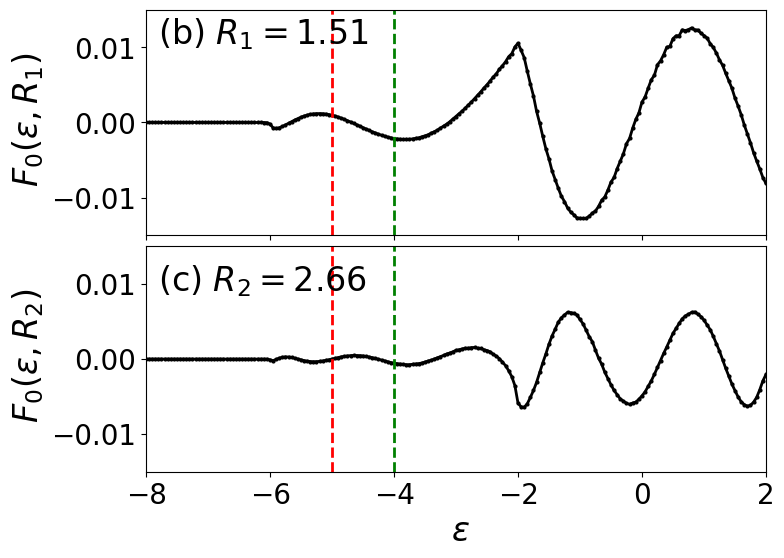}
\caption{(a) Band structure along the high-symmetry points of the Brillouin zone for the 3D tight-binding model (blue) and the free electron model.  (b,c) Integrand of Eq.~\eqref{eq:RKKY_3DTB} for two values of $R=|{\bf R}|/\sqrt{3}$. The dashed lines mark the Fermi energies ($\epsilon_{F}=-5t$ and $\epsilon_{F}=-4t$) used in the RKKY calculations.}
\label{fig:BandStructure_TB_Free}
\end{figure}

Figure \ref{fig:BandStructure_TB_Free}(a) shows $E_{\rm free}(\veck)$ and $E_{\rm TB}(\veck)$ over symmetric paths along the Brillouin zone.   We choose $E_0=-6t$ such that the bottom of the conduction band ($E_{\rm free}(0)=E_{\rm TB}(0)$ at the $\Gamma$ point) coincides in both cases. 

The next step is to calculate the RSGFs (Eq.~\eqref{eq:GF_TB}). We perform the integral over the cubic Brillouin zone ($-\pi \leq k_{x,y,z} \leq \pi$) by constructing a cubic grid of $(N_k)^3$ points. We find good convergence for teh results for $N_k=200-300$ and we use $N_k=400$ in all calculations shown in the paper.

The final step is the energy integral of Eq.\eqref{eq:RKKY_3DTB}. Figure  \ref{fig:BandStructure_TB_Free}(b) shows the integrand of Eq.\eqref{eq:RKKY_3DTB} $F_0(\epsilon,R) \equiv \mbox{Im }  G_{\rm TB}(-{\bf R},\epsilon^{+}) G_{\rm TB}(+{\bf R},\epsilon^{+})$ for ${\bf R}=(R,R,R)$ (impurities at the diagonal of the unit cell)  two values of $R$. Notice that $F_0(\epsilon,R)$ is greatly suppressed for $\epsilon<E_0$ (see Figs.~\ref{fig:BandStructure_TB_Free}(b,c)) and we can safely take $E_{\rm cutoff}=-8t$ as the lower limit of the  energy integral. We typically use the broadening parameter $\delta=0.05t$, which is larger than the energy step in the numerical integration $\delta \epsilon =0.005 t$. Fig.~\ref{fig:BandStructure_TB_Free}  shows that using a smaller value ($\delta=0.01t$) does not change the results significantly.

\begin{figure}[h!]
\centering
\includegraphics[width=0.45\columnwidth]{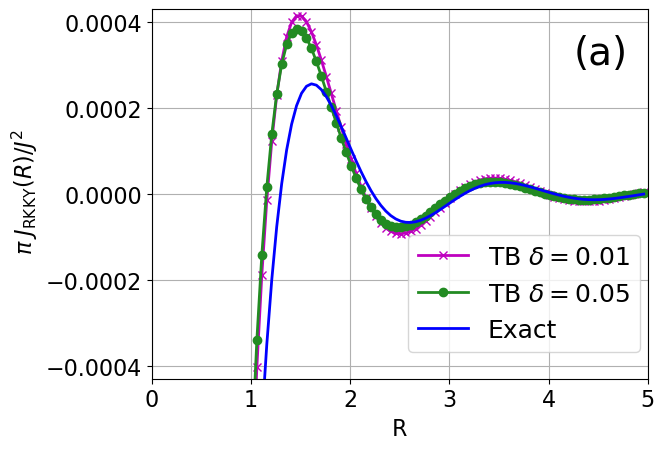}
\includegraphics[width=0.45\columnwidth]{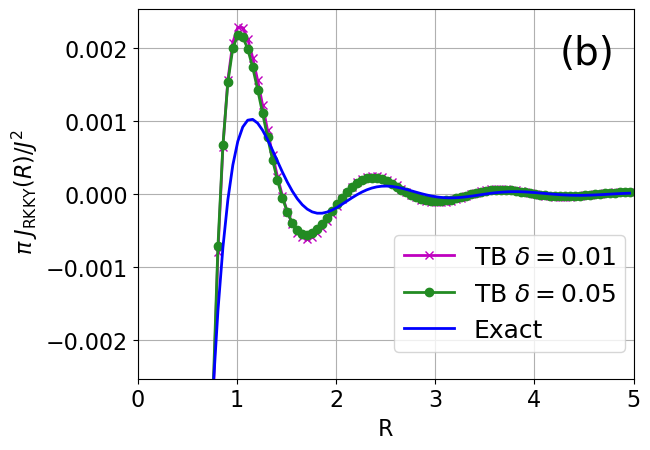}
\caption{ RKKY interaction for the 3D tight-binding model from Eq.~\eqref{eq:RKKY_3DTB} calculated for $\epsilon_{F}=-5t$ (a) and $\epsilon_{F}=-4t$ (b) and different values of the broadening $\delta$. The blue curves corresponds to the analytical expression for the free electron gas (Eq.~\eqref{eq:RKKYanalytic}) for the respective values of $k_F$. }
\label{fig:JRRY_TB_free}
\end{figure}

\begin{figure}[h!]
\centering
\includegraphics[width=0.45\columnwidth]{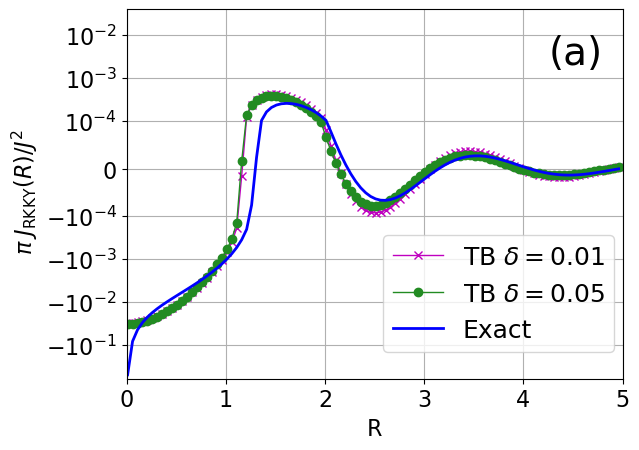}
\includegraphics[width=0.45\columnwidth]{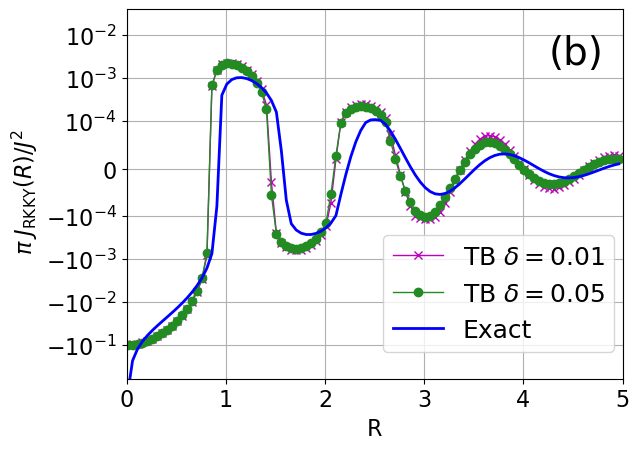}
\caption{Sames as in Fig.~\ref{fig:JRRY_TB_free} in log scale}
\label{fig:JRRY_TB_free_log}
\end{figure}

We calculate $J^{\rm TB}_{\rm RKKY}(R)$  for the two values of $\epsilon_{F}$ ($\epsilon_{F}=-5t$ and $-4t$) shown as dashed lines in Fig.~\ref{fig:BandStructure_TB_Free} and two values of the broadening parameter $\delta$. The results are shown in Figs.~\ref{fig:JRRY_TB_free} and \ref{fig:JRRY_TB_free_log} (in linear and log scales, respectively). In both cases, we compare with the exact result $J^{\rm exact}_{\rm RKKY}(R)$ for the corresponding values of $k_F$ and $|{\bf R}|=\sqrt{3}R$ showing the characteristic $2 k_F R$ oscillations. 

A better agreement is expected for $\epsilon_{F}$ near the bottom of the conduction band (where the dispersion for the tight-binding model is nearly quadratic) and when the distance between the impurities is larger than the lattice spacing ($R \gg 1$). 
In fact, such quantitative agreement is verified, as shown in Figs.~\ref{fig:JRRY_TB_free}(a) and \ref{fig:JRRY_TB_free_log}(a) for $\epsilon_{F}=-5t$ and $R \gtrsim 3$), which validates our choice of the numerical used in the numerical calculations.  For larger values of the Fermi energy (e.g., $\epsilon_{F}=-4t$, shown in Figs.~\ref{fig:JRRY_TB_free}(b) and \ref{fig:JRRY_TB_free_log}(b)), there is a qualitative agreement but the difference in the tight-binding and free dispersions already show up in the results.

\subsection{Cutoff dependency}
\label{sec:cutoff}

In this appendix,  we examine the sensitivity of $J_{ab}(R)$ given by the energy integrals in  Eq.~\eqref{eq:JabIntegral} to the lower energy cutoff $E_{\text{cutoff}}$. In practice, these are calculated using

\begin{equation}
    J_{ab}(R) =-\dfrac{J^{2}}{\pi}\int_{E_{\text{cutoff}}}^{\epsilon_{F}} F_{ab}(R,\epsilon^{+})  \; d\epsilon \; , \label{eq:JabIntegral_cutoff}
\end{equation}

\begin{figure}[h!]
\centering
\includegraphics[width=0.45\columnwidth]{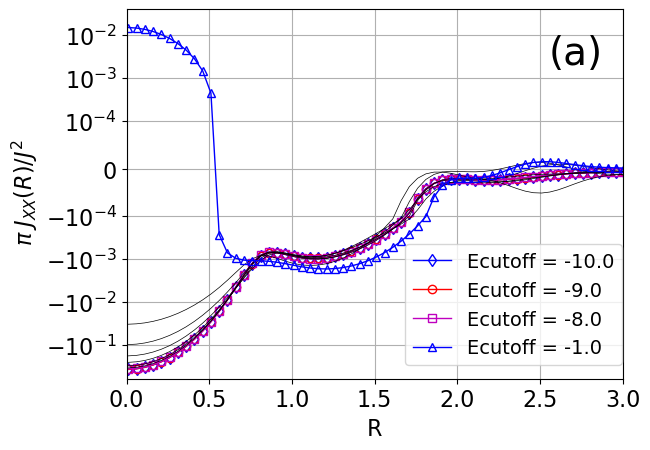}
\includegraphics[width=0.45\columnwidth]{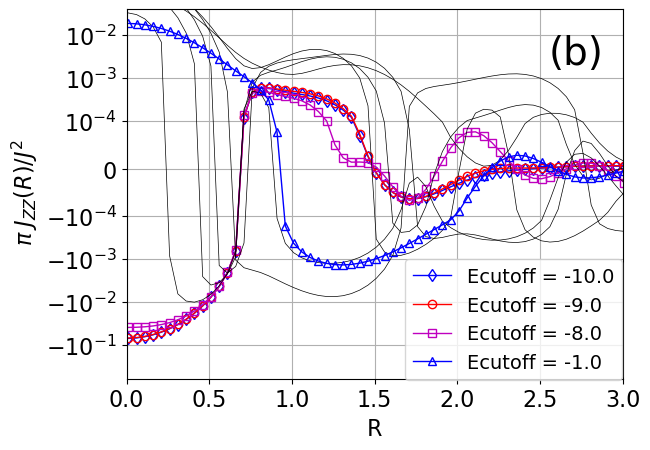}
\caption{Diagonal RKKY matrix elements $J_{XX}(R)$ (a) and $J_{ZZ}(R)$ (b) for different values of the cutoff $E_{\text{cutoff}}$ defined in Eq.~\eqref{eq:JabIntegral_cutoff}. The gray lines represent the intermediate values of $E_{\text{cutoff}}$ not shown in the legends. }
\label{fig:JRRY_cutoff_log}
\end{figure}

Figure \ref{fig:JRRY_cutoff_log} displays $J_{XX}(R)$ and $J_{ZZ}(R)$ calculated for different values of $E_{\text{cutoff}}$. For cutoff values below the lower band edge ($E_{\text{cutoff}} \le -8.0t$), the calculated RKKY exchange profiles exhibit excellent convergence, with negligible variation between $E_{\text{cutoff}} = -8.0t$, $-9.0t$, and $-10.0t$. In contrast, setting a cutoff inside the band structure ($E_{\text{cutoff}} = -1.0t$) truncates the essential occupied spectral weight, inducing spurious phase shifts and artificial oscillations, particularly in $J_{ZZ}(R)$ (Fig.~\ref{fig:JRRY_cutoff_log}(b)). 

As such, it becomes clear that restricting the energy integral to the low-energy regions dominated by the Weyl cones (which, in practice, means $E_{\text{cutoff}} \sim -\Delta_{VHS}/2 \approx -1.0t$) gives nonphysical artifacts, as illustrated by the blue curves in Fig.~\ref{fig:JRRY_cutoff_log}. This confirms that the finite bandwidth of the lattice model naturally cures the ultraviolet divergences inherent to linearized low-energy continuum models, producing physically stable RKKY couplings without requiring ad-hoc regulator functions \cite{Saremi_PhysRevB.76.184430_2007}.

\vspace{1cm}
\bibliographystyle{unsrt}
\bibliography{references}

\end{document}